\documentclass[10pt,letter]{rspublic}

\usepackage{amsmath}
\usepackage{amssymb}
\usepackage{graphicx}
\usepackage{multicol}
\usepackage[numbers]{natbib}
\usepackage{url}

\usepackage{microtype}
\usepackage{bbding}
\usepackage{color,soul}


\usepackage{xr}
\begin{document}

\newcommand{\lt}{\left}
\newcommand{\rt}{\right}
\newcommand{\la}{\langle}
\newcommand{\ra}{\rangle}
\newcommand{\nn}{\nonumber}
\newcommand{\blank}[1]{}
\newcommand{\X}{\tilde{X}}
\newcommand{\phit}{\tilde{\phi}}
\newcommand{\La}{\Lambda}
\newcommand{\ad}{a^{\dagger}}
\newcommand{\ep}{\varepsilon}
\newmuskip\pFqmuskip

\newcommand*\pFq[6][8]{%
  \begingroup 
  \pFqmuskip=#1mu\relax
  \mathcode`\,=\string"8000
  \begingroup\lccode`\~=`\,
  \lowercase{\endgroup\let~}\pFqcomma
  {}_{#2}F_{#3}{\left[\genfrac..{0pt}{}{#4}{#5};#6\right]}%
  \endgroup
}
\newcommand{\pFqcomma}{\mskip\pFqmuskip}

\title[Neutrality and the role of rare variants]{Inferring processes of cultural transmission: the critical role of rare variants in distinguishing neutrality from novelty biases}
   \author[JP O'Dwyer, A Kandler]{James P. O'Dwyer$^1$ \& Anne Kandler$^2$}
   \affiliation{Department of Plant Biology, University of Illinois, Urbana IL 61801 USA$^1$ \\ Max Planck Institute for Evolutionary Anthropology, Department of Human Behavior, Ecology and Culture, Leipzig, Germany$^2$}
   \label{firstpage}
   \maketitle

\setcounter{section}{0}
{\bf Abstract.}
Neutral evolution assumes that there are no selective forces distinguishing different variants in a population. Despite this striking assumption, many recent studies have sought to assess whether neutrality can provide a good description of different episodes of cultural change. One approach has been to test whether neutral predictions are consistent with observed progeny distributions, recording the number of variants that have produced a given number of new instances within a specified time interval: a classic example is the distribution of baby names. Using an overlapping generations model we show that these distributions consist of two phases: a power law phase with a constant exponent of -3/2, followed by an exponential cut-off for variants with very large numbers of progeny. Maximum likelihood estimations of the model parameters provide a direct way to establish whether observed empirical patterns are consistent with neutral evolution. We apply our approach to a complete data set of baby names from Australia. Crucially we show that analyses based on only the most popular variants, as is often the case in studies of cultural evolution, can provide misleading evidence for underlying transmission hypotheses. While neutrality provides a plausible description of progeny distributions of abundant variants, rare variants deviate from neutrality. Further, we develop a simulation framework that allows for the detection of alternative cultural transmission processes. We show that anti-novelty bias is able to replicate the complete progeny distribution of the Australian data set.

Keywords: Cultural transmission, neutral evolution, pro-novelty bias, anti-novelty bias, progeny distribution, power law

\pagebreak 
\section{Introduction}

Most theoretical modelling frameworks to cultural evolution make the simplifying assumption that innovations are the product of erroneous cultural transmission resulting in the introduction of cultural variants not previously seen in the population at low abundances \citep[e.g.][]{cavalli1981cultural,boyd1988culture}).
But regardless of the mechanisms underlying the occurrence of any particular innovation, its subsequent fate (i.e. whether it goes extinct immediately or is able to spread through the population and reach a certain degree of visibility) provides a window into the processes of cultural transmission present in the population. For example, the `persistence' of a large number of innovations might point to population-level preferences for novel or rare variants. As a large number of such cultural transmission hypotheses have been proposed in the literature \citep[see e.g.][]{laland2004social}, the question whether we can develop systematic approaches to distinguish between different transmission hypotheses using aggregated population-level data has gained importance. 

Seminal work by Bentley and colleagues \citep[e.g.][]{bentley2004random,hahn2003drift,herzog2004random} on this topic has focused on distinguishing broadly between neutral and non-neutral cultural transmission processes. Neutral models of cultural transmission make the assumption that there are no selective differences between variants, so that the dynamics of a new variant are not biased either towards proliferation or extinction.  This hypothesis results in a  particular kind of stochastic dynamics, known as drift. In balancing the utility and availability of cultural data, the studies mentioned above identified the progeny distribution as a way to distinguish the neutral hypothesis from others.  The progeny distribution logs the abundances of cultural variant types which produce $k$ new individuals over a fixed period of time. 
Bentley and colleagues have estimated the form of the neutral progeny distribution through simulation techniques~ \citep[e.g.][]{bentley2004random,hahn2003drift,premo2014cultural}, concluding that the progeny distribution takes the form of a power law. The exponent of this power law has been fitted as a function that depends on innovation rate and total population size. The theoretical predictions have been compared against empirical data for the choice of baby names, US patents and their citations or pottery motifs, and these analyses provided support for the neutral hypothesis~\citep{bentley2004random,hahn2003drift}. Despite this progress, an analytical expression for the neutral progeny distribution has been lacking so far, which limited further developments in understanding whether observed distributions are consistent with neutrality, or demand non-neutral explanations.

In this manuscript we derive the first analytical representation of the neutral progeny distribution for large time intervals, using a neutral model where variants are not constrained to produce at discrete time points, known as an overlapping generations model. We show that the neutral progeny distribution consists of two phases. For small numbers of progeny there is a power law phase. This is broadly consistent with the fits to earlier numerical simulations, but here we find that this power law has a fixed, universally-applicable exponent of -3/2. Following this power law phase, for large enough numbers of progeny there is eventually an exponential drop-off  in this distribution. The onset of the exponential decline depends on the innovation rate: the larger the rate, the earlier the onset.  The analytical representation of the progeny distribution allows for maximum-likelihood estimations of the model parameter and therefore provides a direct way of parametrizing neutral models using cultural data, and of subsequently evaluating the consistency between observed data and the neutral hypothesis. Importantly we establish that analyses based on only the most popular variants, as is often the case in studies of cultural evolution, can provide misleading evidence for neutral evolution. 

Further, we show that the progeny distribution represents a statistic that is able to detect alternative cultural transmission hypotheses, in particular bias for or against novelty, and therefore is potentially capable of distinguishing between different processes of cultural transmission based on population-level data. For that we develop a simulation procedure which includes pro- and anti-novelty bias. Anti-novelty bias is characterized as the preference for variants which have been present in the population for a long time (i.e. innovations possess an intrinsic disadvantage) while pro-novelty bias describes the preference for `young' variant types that have only recently been introduced into the cultural system (i.e. innovations possess an intrinsic advantage). In general we find that the progeny distribution reacts sensitively to those changes in the transmission process.
Related results have been found by~\citep{mesoudi2009random} who concluded that strong frequency-dependent biases alter the shape of the progeny distribution. 
They also note that some transmission biases will generate population-level predictions indistinguishable from neutral predictions. 

Following~\citep{hahn2003drift}, we apply our framework to an Australian data set recording the first names of newborns (The code of the simulation framework can be downloaded under \url{ https://github.com/odwyer-lab/neutral_progeny_distribution}.). We demonstrate the importance of rare variants for reliable inference of processes of cultural evolution from aggregated population-level data in form of progeny distributions. While the temporal dynamics of abundant names are consistent with neutrality, the analysis based on the complete distribution, including popular and rare names, provides evidence against neutral evolution. 
This means that progeny distributions generate reliable inferences only in situations where the complete data set is available. 
We find that anti-novelty bias is able to replicate the complete progeny distribution of the considered Australian baby name data. 

\section{Neutral Theory and Innovation}\label{sec:neutral}

Neutral models have provided basic null models in fields stretching from population genetics~\citep{Kimura1968} and ecology~\citep{Hubbell1979,Hubbell2001, Chave2004,Rosindell2011,Odwyer2012d}, to cultural evolution and the social sciences~\citep[e.g.][]{bentley2004random,neiman1995stylistic,Shennan2001,kohler2004vessels}. At the core of all varieties of neutral theory is a group of competing variants, and the assumption that selective differences between these variants are  absent. In addition, most neutral models contain the possibility for innovation, i.e. the introduction of entirely new variants into the system. The most common approach to modeling an innovation event is to assume that with some rate
a parent individual will produce an offspring of a new type instead of an offspring of the same parental type. 
This new variant then undergoes the same dynamics as all extant variants.

The assumptions of neutrality are often at odds with the vast stores of knowledge biologists and anthropologists have accumulated for natural and social systems. For example, we know that even closely-related biological species differ in their phenotype, and we might expect that these differences are important for predicting and understanding the properties of ecological communities. And yet despite this obvious roadblock, neutral models in ecology have had some considerable success in predicting patterns of biodiversity observed at a single snapshot in time~\citep{Volkov2003,Etienne2005,Etienne2005b,Volkov2007,Etienne2007, Chisholm2009,Condit2002,Chave2002b,Rosindell2008,Rosindell2009,Odwyer2010,Vellend2010}. 
The same is true for cultural evolution where humans are generally not thought of as making decisions at random. Neutrality would imply that individuals do not possess any preferences for existing cultural variants, nor does the adoption of a particular cultural variant provide an evolutionary advantage over the adoption of a different variant. While these inherent assumptions are likely to be violated in the cultural context (for detailed discussions see \citep[e.g.][]{neiman1995stylistic,Shennan2001,steele2010ceramic})
population-level patterns of various observed episodes of cultural change nevertheless resemble the ones expected under neutrality \citep[e.g.][]{bentley2004random,neiman1995stylistic,bentley2007regular}. 

Statistical tests of neutral theory often focus on static patterns of diversity, observed at one moment in time, such as the balance of rare and dominant species in a population. It has been shown that neutral steady-state predictions for the distribution of species abundances often closely match observed distributions. In contrast, neutral theories in ecology have had less success in predicting the dynamics of diversity, from decadal-scale species abundance fluctuations to geological ages of species~\citep{Leigh2007,Wang2013,chisholm2014ages,Fung2015,odwyer2015phylo}. 
Similarly, recent work in cultural evolution has pointed to the 
importance of analyzing temporal patterns of change as opposed to static measures of cultural diversity \citep[e.g.][]{mcelreath2005applying,hoppitt2010detecting,kandler2013non,kandler2015generative} and to the influence of aggregation processes particularly in archaeological case studies \citep{premo2014cultural}
when testing for departures from neutrality. 
At the very least, these discrepancies bring to light the importance of what statistics are chosen to test a hypothesis like neutral evolution. 
In this light, a recent study \citep{sindi2016culturomics} analysed the patterns of frequency change, in particular the kurtosis of the distribution of changes over time, of stable words in the {\it Google Ngram} data base. Interestingly, this approach identified words under selection: kurtosis values close to zero signaled neutrality while deviations from zero were indicative of selection.

In this paper we apply ecological neutral theory to cultural data. We use a model that allows for overlapping generations, an appropriate assumption when analyzing distributions of cultural variants, and for an analytical representation of the progeny distribution. In the following we provide a brief review of the characteristics of this model.

\subsection{Neutral Theory in Ecology}\label{sec:neutral_ecology}

It is assumed that the temporal dynamics of species are governed by reproduction and competition, occurring in continuous time with a given set of rates. 
The full, interacting version of this model can be described by stochastic Lotka-Volterra systems 
(either with symmetric, pairwise competition between species where the strength of the competition is controlled by the constant $\alpha$, or any related constraint on population size). Solving for the dynamics of these systems is, however, analytically intractable but a solvable mean field approximation has been found. This approximation is based on treating each species as interacting with the average state of all other species, rather than the specific configuration of abundances at any given moment in time~\citep{Volkov2003,Volkov2007,odwyer2014redqueen}.  In the limit of a large number of species this approach states that the correlation between the abundances of any two species is assumed to be small. In other words, the abundances of extant species are assumed to evolve independently of each other. 
Importantly, the resulting mean field description collapses non-linear rates of competitive interaction into an increased, linear mortality rate for each species.
This approximation of the overlapping generations neutral model is also known as the `non-zero-sum' or NZS approximation referring to the fact that the total population size may fluctuate over time, i.e. births and deaths do not sum to zero. It has been shown that this approach provides only a good approximation in populations with a large number of species, but in a less diverse population, where a handful of species are dominant, the mean field approximation is no longer a meaningful description.

\begin{figure}[htb]
\begin{center}
\includegraphics[width=9cm]{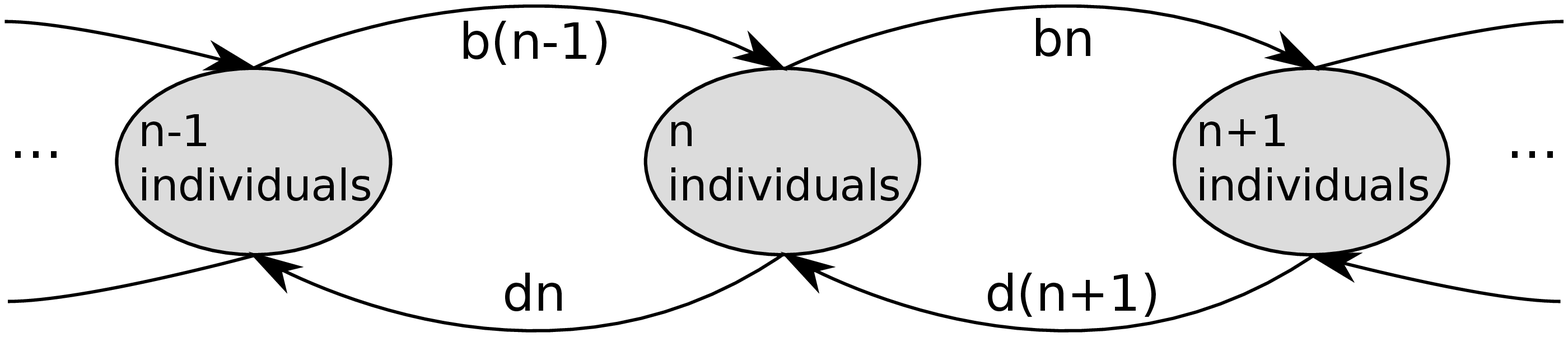}
\end{center}
\caption{Schematic representation of the birth-death dynamic described in Eq.~\eqref{eq:master} where the variables $b$ and $d$ stand for birth and death rates, respectively.}\label{fig:model}
\end{figure}

In the mean field approximation, each species takes an independent, random walk, based on a linear stochastic process.  
Mathematically, this is described by a linear master equation for the probability $P(n|t)$ that a species has abundance $n$ conditioned on its age (i.e. time since introduction into the system)
\begin{equation}\label{eq:master}
\frac{dP}{dt} = b(n-1)P(n-1|t)-bnP(n|t)-dnP(n|t)+d(n+1)P(n+1|t).
\end{equation}
Here, $t$ is species age, and for so-called `point' speciation (where new species always have an abundance of 1) the initial condition is $P(n|0) = \delta_{n,1}$ (see Fig.~\ref{fig:model} for a schematic representation of the model dynamic). 

The value $d$, which is always strictly larger than the birth rate, $b$, is a combination of intrinsic mortality and the effect of competition arising from all other species. For the point speciation process, this linear master equation has the time-dependent solution
\begin{align}\label{eq:solution_master}
P(n|t) = e^{(b-d)t}\frac{\lt(\frac{b}{d-b}(1-e^{(b-d)t}\rt)^{n-1}}{\lt(1+\frac{b}{d-b}(1-e^{(b-d)t}\rt)^{n+1}}.
\end{align}
For a more general initial condition, there is a correspondingly more general solution (see Section~S2 in the supplementary material for a detailed mathematical derivations of these results).

Eq.~\eqref{eq:solution_master} describes the temporal dynamics of a single species, from its introduction into the system to (guaranteed) eventual extinction. Under the additional assumption that in steady state, the rate of appearance of new species in a population of size $J$
is given by $\nu J$, it can be shown that the expected species abundance distribution (i.e. the number of species with abundance $k$) takes the form of a log series distribution
\begin{align}\label{eq:speciesAbundanceNSZ}
\langle S(k)\rangle 
& \simeq \nu J \int_0^{\infty} P(k|t)dt
\simeq \frac{\theta}{k} \lt(1-\frac{\theta}{J}\rt)^k
\end{align}
where $\theta=\lt(1-\frac{b}{d}\rt) J$ stands for the `fundamental biodiversity number'. Finally, there is a constraint relating speciation rate $\nu$ to $b$ and $d$ rooted in the mean field approximation.  The parameter $d$ is an effective parameter arising from the influence of the rest of the population and therefore the  per capita speciation rate $\nu$ is constrained to be related to these rates as
\begin{equation}
\nu = {d}-{b}.
\end{equation}

Summarizing, Eq.~\eqref{eq:solution_master} gives a complete description of the non-spatial, NZS model that provides a good approximation to various neutral predictions in ecology when diversity is high~\citep{Volkov2003,Volkov2007,chisholm2014ages,odwyer2014redqueen,Etienne2009}. 

To ensure consistent notation across different scientific disciplines we will refer in the following to species as variants, to individuals as instances and to speciation as innovation. Further, birth and death rates describe the rates at which a cultural variant generates or looses an instance, respectively (see Fig.~\ref{fig:model}). 

\subsection{Neutral Theory in Cultural Evolution}

Neutral theory in cultural evolution has been mainly modelled using the Wright Fisher infinitely many allele model (see e.g. \citep{ewens2012mathematical} for a review of the mathematical properties, \citep{neiman1995stylistic} for its introduction to cultural evolution as well as \citep[e.g.][]{bentley2004random,Shennan2001,kohler2004vessels,steele2010ceramic} for further applications to cultural case studies).   
In general, this framework assumes that the composition of the population of instances of cultural variants at time $t$ is derived by sampling with replacement from the population of instances at time $t-1$ resulting in non-overlapping generations.  
We provide in Section~S1 of the supplementary material a brief review of the mathematical characteristics of this model.

\section{The Neutral Progeny Distribution}\label{sec:neutralprogdbn}

Data sets describing the accumulated appearances of cultural variants within a specific time interval, like the choice of baby names in human populations, have typically been summarized by the progeny distribution. 
This distribution logs the frequency of cultural variants with a total of $k$ progeny, taken over a given, fixed duration, $T$.  In part, this choice of distribution is pragmatic; data for baby names registered at birth are often more complete and more readily available than full censuses of names in a population, which would provide the analogue of a species abundance distribution given in Eq. \eqref{eq:speciesAbundanceNSZ}.
Additionally, the progeny distribution contains a temporal element, as in general the distribution will change with the duration, $T$, that the progeny counts are taken over. Finally, the progeny distribution is particularly useful for populations where the effective population size of reproducing individuals may be much smaller than the total population. The distribution directly probes the dynamics of transmission of cultural variants, whereas the species abundance distribution may be much more sensitive to the details of the age structure in the population.

In this section we derive an analytical representation of the progeny distribution based on the overlapping generation NZS model for large, well-mixed populations. We show, in agreement with earlier work, that neutral theory generates a power law progeny distribution but with a constant exponent of -3/2, (i.e. the power law exponent does {\it not} depend on innovation rate or population). 
The power law is followed by an exponential cut-off, whereby the onset of this cut-off depends in the innovation rate. 
Further, we provide a method for identifying  maximum likelihood neutral parameters.

\subsection{Analytical results}\label{sec:nzsprog}

Using the NZS approximation the progeny distribution at late times $T$, i.e. under the assumption that sufficient time has passed so that the distribution reached stationarity, can be derived as 
\begin{equation}\label{eq:late-time}
q(k)= (-1)^{k-1}{{\frac{1}{2}}\choose{k} }\frac{2d}{b+d}\lt(\frac{4bd}{(b+d)^2}\rt)^{k-1}
\end{equation}
where $b$ and $d$ stand for the birth and death rates of the variants (see Section~S3 in the supplementary material for a detailed derivation) and the term 
${\frac{1}{2}}\choose{k}$ is defined by
\[
{{\frac{1}{2}}\choose{k}}={{2k}\choose{k}}\frac{-1^{k+1}}{2^{2k}(2k-1)}.
\]
The function $q(k)$ describes the frequency of cultural variants which generated exactly $k$ instances, including its innovation event, within a time interval of length $T$.
Eq.~\eqref{eq:late-time} is valid only in the large $T$ limit, but in Section~S3 of the supplementary material we also provide additional results for moments and generating functions of this distribution for arbitrary  durations, $T$. The corresponding cumulative distribution (i.e. the fraction of variants with greater than or equal to $k$ cultural variants generated within a time interval of length $T$) is given by
\begin{equation}\label{eq:late-time-cumulative}
\hspace*{-0.25cm}P(K\ge k) =  (-1)^{k-1}\frac{b+d}{2b}\lt(\frac{4bd}{(b+d)^2}\rt)^k{{\frac{1}{2}}\choose{k} }\pFq{2}{1}{1,(-1/2+k)}{1+k}{\frac{4bd}{(b+d)^2}}
\end{equation}
with $\pFq{2}{1}{\cdot ,\cdot}{\cdot}{\cdot}$ representing the Gaussian hypergeometric function (see Section~S3 in the supplementary material for a detailed derivation).

\begin{figure}[htb]
\begin{center}
\includegraphics[width=9cm]{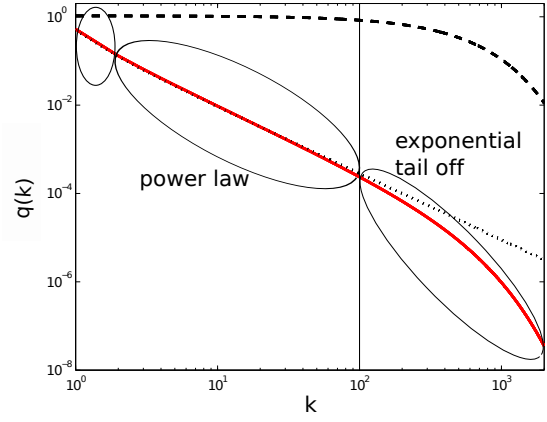}
\end{center}
\caption{Illustration of the shape of the progeny distribution \eqref{eq:late-time} with $b=1$ and $d=1.1$ (red line). The dashed line shows the term $(-1)^{k-1} {{\frac{1}{2}}\choose{k} }$ and the dotted line the term $\frac{2d}{b+d}\lt(\frac{4bd}{(b+d)^2}\rt)^{k}$. The black vertical line at $(b/(d-b))^2$ indicates the approximation of the transition point from the power law behavior to the exponential decay.}\label{fig:illustration}
\end{figure}

Interestingly, the distribution $q(k)$ fragments into two parts: one describes a power law and the other an exponential decay (see dotted and dashed lines in Fig. \ref{fig:illustration}).  
For large enough values of $k$ the first terms of Eq. \eqref{eq:late-time} can be approximated by
\begin{align}
(-1)^{k-1} {{\frac{1}{2}}\choose{k} }& \simeq (-1)^{k-1} \frac{(-1)^{k}}{\Gamma\lt(-\frac{1}{2}\rt)k^{\frac{3}{2}}}\nn
=\frac{1}{2\sqrt{\pi}k^{\frac{3}{2}}}
\end{align}
which determines a power law with the exponent $-3/2$. However, at 
approximately $k\sim(b/(d-b))^2=\lt({b}/{\nu}\rt)^2$ the exponential decay starts dominating the distribution (see red line in Fig. \ref{fig:illustration}).
In summary, the neutral progeny distribution tends towards a power law with an universally-applicable exponent of $-3/2$ (i.e. the exponent does not, as previously suggested, depend on the parameters of the neutral model) but shows an exponential cut-off at approximately $k\sim(b/(d-b))^2=\lt({b}/{\nu}\rt)^2$. The larger the innovation rate, $\nu/d$, the smaller the values of $k$ for which exponential decay dominates the progeny distribution. 

\subsection{Maximum Likelihood Parameters}

To fit the progeny distribution given in Eq.~\eqref{eq:late-time} to empirical data
we derive the maximum likelihood estimate of the ratio $\eta=d/b$ (as we show below that the shape of the progeny distribution depends only on the ratio of the death and birth rate).       

The log likelihood of observing a given set of $S$ cultural variants with abundances $\{k_i\}$ at late times is given by
\begin{equation*}
L = \sum_{i=1}^S    \log\lt[\frac{2d}{b+d}\lt(\frac{4bd}{(b+d)^2}\rt)^{k_i-1} (-1)^{k_i-1} {{\frac{1}{2}}\choose{k_i} }\rt]
\end{equation*}
which can be rewritten as
\begin{equation*}
L =\sum_{i=1}^S   \log\lt[\frac{2\eta}{1+\eta}\lt(\frac{4}{\frac{1}{\eta}+2+\eta}\rt)^{k_i-1} (-1)^{k_i-1} {{\frac{1}{2}}\choose{k_i} }\rt]
\end{equation*}
by using the relation
$\eta= \frac{d}{b}$. 
Maximizing this log likelihood with respect to parameter $\eta$ provides the following point estimate
\begin{align}\label{eq:maxlik_eta}
\eta = \frac{K_{\text{total}}}{K_{\text{total}}-S}
\end{align}
where $K_{\text{total}}$ is total number of instances observed in the data and $S$ is the total number of variant (a detailed derivation can be found in Section~S4 of the supplementary material).

\subsection{Comparison of Analytical Approximations with Simulations}\label{sec:sims}

In this section we ensure the validity of our approximations (in particular Eqs. \eqref{eq:late-time} and \eqref{eq:late-time-cumulative}) by comparing analytical and numerical results.
To do so  we simulate the full, non-linear model with overlapping generations.  
In detail, we generate the temporal frequency behavior of a group of competing variants via the Gillespie algorithm and compute the resulting progeny distribution after a long time interval. 

We use stochastic Lotka-Volterra systems, where variant $i$ with current abundance $n_i$ 
will undergo birth and death processes as well as be involved in competitive interactions with other variants. 
New variants are introduced at a rate $\nu J$ ($J$ describes the total population size) with initial abundance $1$, and are considered as an error in the birth process. Therefore the effective per capita birth rate is given by $b_0-\nu$.  The rates of these processes for variant $i$ are as follows
\begin{align}\label{eq:summary_processes_rates}
\hspace{0.1cm}\textrm{process} \hspace{1cm}& \hspace{1cm} \textrm{description} \hspace{0.5cm} & \hspace{0.1cm} \textrm{rate}\nn\\
\hspace{0.1cm} n_i \rightarrow n_i +1\hspace{1cm} &\hspace{1cm} \textrm{birth} \hspace{0.5cm}&\hspace{0.1cm} (b_0-\nu)n_i\nn\\
\hspace{0.1cm}n_i \rightarrow n_i -1\hspace{1cm} & \hspace{1cm}\textrm{intrinsic mortality}\hspace{0.5cm} & \hspace{0.1cm} d_0n_i\nn\\
\hspace{0.1cm}n_i \rightarrow n_i -1\hspace{1cm} &\hspace{1cm} \textrm{competition}\hspace{0.5cm} &\hspace{0.1cm} \alpha n_i \sum_{\forall\,j} n_j\nn\\
\hspace{0.1cm}0 \rightarrow 1 \hspace{1cm} &\hspace{1cm} \textrm{speciation}\hspace{0.5cm} &\hspace{0.1cm} \nu\sum_{\forall\,j} n_j
\end{align}
where the labels $i$ and $j$ refer to the extant variants in the system at any given point in time, and the sums are taken over all variants, including variant type $i$.
The simulation of this population is based on the well-known Gillespie algorithm~\citep{gillespie1976general}. We provide a detailed description of the simulation procedure in Section~S5 of the supplementary material. The code used is available under\\  \url{https://github.com/odwyer-lab/neutral_progeny_distribution}. 

\begin{figure}[htb]
\setlength{\unitlength}{0.1\textwidth}
\begin{center}
\includegraphics[angle=0,scale=0.28,trim= 0 0 0 0,clip=TRUE]{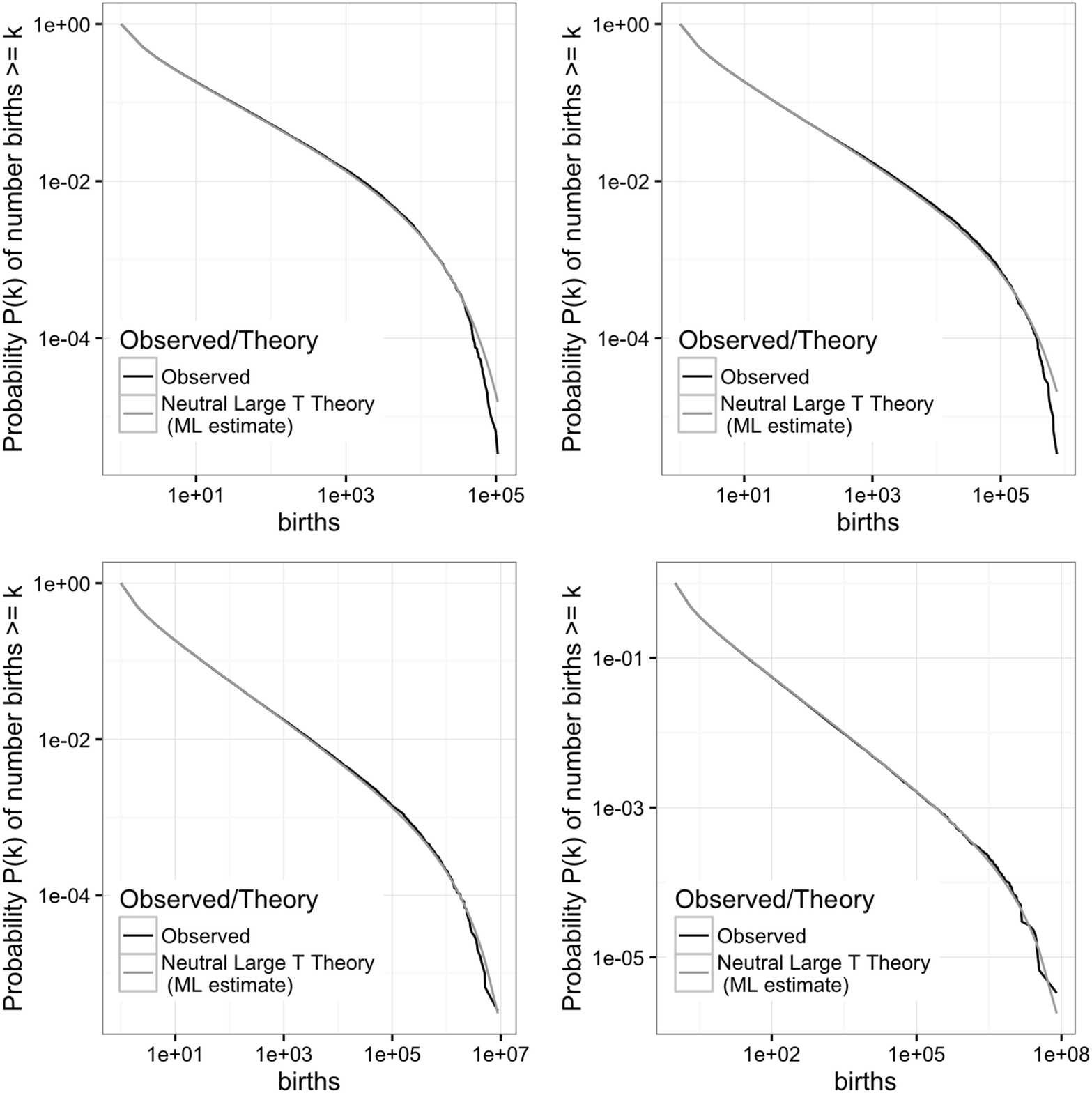}
\end{center}
\caption{Cumulative progeny distributions for long time intervals 
determined by simulated, neutral populations based on competitive Lotka-Volterra interactions with overlapping generations (black lines) match their mean field, non-zero sum approximations \eqref{eq:late-time-cumulative} (grey lines). We consider parameter values (top, left figure) $J=300$, $\nu=0.01$, (top, right figure) $J=1000$, $\nu=0.003$, (bottom, left figure) $J=3000$, $\nu=0.01$, (bottom, right figure) $J=10000$, $\nu=0.0003$. In all cases, the product $J\nu$ has been chosen to be $\simeq3$, so that the mean field regime is reached, and in each case the simulated progeny distribution was logged for $T=100000$ generations.}\label{fig:speciesrichness}
\end{figure}

Fig.~\ref{fig:speciesrichness} illustrates that the simulated cumulative progeny distributions based on competitive Lotka-Volterra interactions
(black dots) coincide their analytical counterparts given by Eq. \eqref{eq:late-time-cumulative} (red lines) for long time intervals and various values of $\nu$ and $J$. 
In summary , Eq. \eqref{eq:late-time-cumulative} (and consequently Eq. \eqref{eq:late-time}) provides an accurate description of the neutral predictions for a model with symmetric, competitive interactions and overlapping generations. 

\section{Novelty Biases}\label{sec:anb}

So far we assumed that there are no selective differences between the extant variants in the population. 
In this section we generalise our framework to include selection for and against novel cultural variants (denoted as pro-novelty bias and anti-novelty bias, respectively) and explore the consequences of these selection biases on the shape of the progeny distribution.  

In general, pro-novelty selection favours `young' variants, i.e. variants that have been invented recently. In contrast, anti-novelty selection disadvantages `young' variants and therefore favors the persistence of established cultural variants over a long time period.  In cultural evolution, pro-novelty selection has been associated with fashion trends~\citep{kandler2015generative,acerbi2012logic}, i.e. the phenomenon where some cultural variants rapidly increase in frequency but also quickly fade away again after other variants have become fashionable. An ecological analog to pro-novelty bias is the red queen effect which is well-explored in the literature (e.g.~\citep{odwyer2014redqueen}). While the red queen effect is typically thought to arise from the accumulation of selectively advantageous traits over time, the emergent effect is an advantage for new species.






\subsection{Pro-novelty bias}

We model pro-novelty bias following earlier ecological theory developed in the context of the red queen hypothesis~\citep{odwyer2014redqueen}.  The only change relative to the simulation described in Section~\ref{sec:neutralprogdbn}(\ref{sec:sims}) is the form of the competition between older and younger variants.  The rate $\alpha_{ij}$ now encodes the competitive effect of species $j$ on species $i$, and depends on innovation times (i.e. the ages of the variants) $\tau_j$ and $\tau_i$
\begin{align}
\alpha_{ij} & = \alpha(1-\ep_0 )\ \ \textrm{for $\tau_j>\tau_i$}, \nn\\
\alpha_{ii} & = \alpha,\nn\\
\alpha_{ij} & = \alpha(1+\ep_0)\ \ \textrm{for $\tau_i>\tau_j$}. \label{eq:proglandscape}
\end{align}
This means we assume that new variants have the same competitive advantage over all extant variants and each variant interacts with three groups: newer, more advantageous variants, conspecifics and older, less advantageous variants~\citep{odwyer2014redqueen}. 
The coefficient $\alpha$ characterizes the strength of competition, while $\ep_0$ is a constant between zero and one that introduces asymmetry in the competitive interactions. 

\begin{figure}[htb]
\begin{center}
\includegraphics[width=0.9\textwidth]{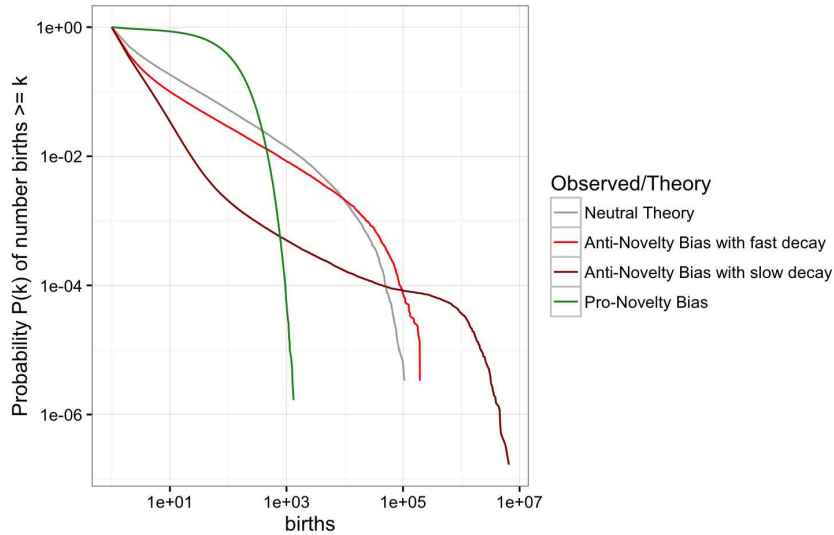}
\end{center}
\caption{Cumulative progeny distributions generated by neutral theory (grey line), pro-novelty bias (green line) and anti-novelty bias (light red and dark red lines) for $J=300$ and $\nu=0.01$. For pro- and anti-novelty bias the asymmetry parameter is set to  $\ep_0=-1$. For anti-novelty bias the decrease of the competitive difference is chosen to be $\lambda=0.3$ (dark red) and $\lambda=3$ (light red line). In each case the simulated progeny distribution was logged for $T=100000$ generations. }\label{fig:comp_neutral_novelty}
\end{figure}

Fig. \ref{fig:comp_neutral_novelty} shows the progeny distributions generated by neutral theory (grey line), pro-novelty selection (green line) for the parameter constellation $J=300$, $\nu=0.01$ and $\ep_0=1$. It is obvious that pro-novelty bias leads to a higher number of variants with small and intermediate abundances and a lower number of variants with very high abundances. As expected, pro-novelty bias reduces the number of singletons, i.e. innovations that have never been transmitted and therefore remained at abundance one.

\subsection{Anti-novelty bias}

Modelling anti-novelty bias in a plausible way is not as straightforward as pro-novelty bias.  If we were to take the competition coefficients given in \eqref{eq:proglandscape} and flip the signs, it is highly likely that, for realistic population sizes, we will end up with one, eternal, old variant, and all other variants that enter the system are driven to extinction over a relatively short time frame.  While we would expect that anti-novelty bias should promote the persistence of older variants, with a strict competitive advantage of all older variants over all newer variants, these results are too extreme.  

We therefore introduce the following rates $\alpha_{ij}$ for the competitive effect of variant $j$ on variant $i$, which again depend on innovation times $\tau_j$ and $\tau_i$ but contain an additional exponential decay factor
\begin{align}
\alpha_{ij} & = \alpha(1-\ep_0e^{-\lambda\tau_j} )\ \ \textrm{for $\tau_j<\tau_i$}, \nn\\
\alpha_{ii} & = \alpha,\nn\\
\alpha_{ij} & = \alpha(1+\ep_0e^{-\lambda\tau_i})\ \ \textrm{for $\tau_i<\tau_j$}. \label{eq:antiproglandscape}
\end{align}
where now we consider $\ep_0<0$, and $\lambda>0$. The effect of $\lambda$ is that  as a variant ages, competitive differences decrease and they begin to interact more and more symmetrically. This approach allows for the persistence of multiple, older variants, because once a type has survived for a time larger than $1/\lambda$, it interacts almost neutrally with all other established variants.

Fig. \ref{fig:comp_neutral_novelty} shows the progeny distributions generated by neutral theory (grey line) and anti-novelty selection (light red and dark red lines) for the parameter constellation $J=300$, $\nu=0.01$, $\ep_0=-1$, $\lambda=0.3$ (dark red line) and $\lambda=3$ (light red line). Anti-novelty bias leads to a lower number of variants with small and intermediate abundances and a higher number of variants with very high abundances. As expected, anti-novelty bias generates a large number of singletons. Further, the slower the decay of the bias, i.e the smaller $\lambda$, the more pronounced are the differences between neutral evolution and anti-novelty selection.

\section{Empirical Analysis for Baby Names}
\label{sec:baby_names}

Starting with the work by~\citep{hahn2003drift} data on the choice of baby names have been widely analyzed in the literature using a variety of frameworks. For example,~\citep{barucca2015cross} analysed the spatial clustering patterns in regard to baby names choice between US states~\citep[see also][]{bentley2012accelerated} or~\citep{acerbi2014biases} used turnover rates to detect transmission biases in US baby names. Further,~\citep{kessler2012you} aimed at disentangling stochastic and deterministic influences on the choice of first names. They suggested that the individual trajectories of name frequencies can be replicated by a deterministic dynamic governed by memory and delay processes.

Here we apply our methodology to two data sets drawn from the state of South Australia, consisting of {\it all} boys' and girls' names registered from 1944 to 2013, respectively, and explore the conclusions about the evolutionary process that can be drawn form it. 
These data sets are included in Section~S6 of the supplementary material together with and a general description and a justification of the application of the mean field approach. 

\subsection{South Australia Baby Names, Neutrality, and Novelty Disadvantage}
\label{subsec:sa_analysis}

First, we calculate the maximum likelihood estimate \eqref{eq:maxlik_eta} of the neutral innovation rate, i.e. the rate that most closely explains the observed progeny distributions computed over the full time span of the data sets. We obtain 
\begin{align}
\left.\frac{\nu}{d}\right|_{\textrm{girls}}=0.05 \qquad\text{and}\qquad
\left.\frac{\nu}{d}\right|_{\textrm{boys}}  = 0.03
\end{align}
indicating a higher tendency for choosing a unique name for newborn girls than for newborn boys. 

\begin{figure}[htb]
\setlength{\unitlength}{0.1\textwidth}
\begin{center}
{\includegraphics[angle=0,scale=0.23,trim= 0 0 0 0,clip=TRUE]{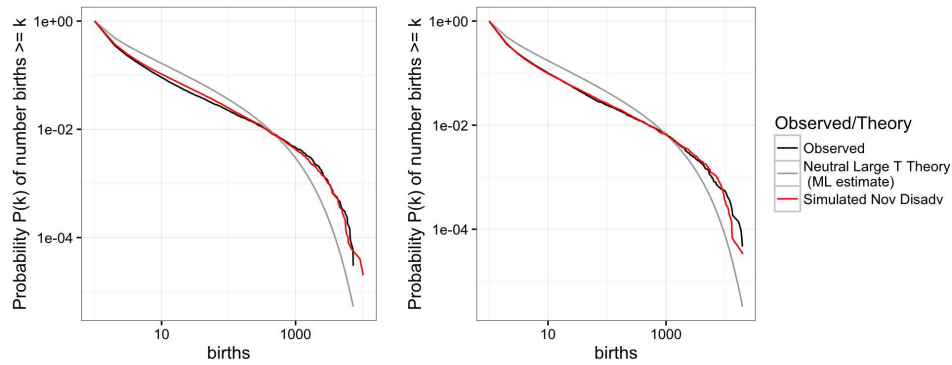}}
\end{center}
\caption{Cumulative progeny distributions for South Australian girls names (left figure) and boys names (right figure). Black lines: empirical distributions over a seventy year span, grey lines: neutral progeny distribution with maximum likelihood parameters, red lines: progeny distribution for novelty disadvantage.} 
\label{fig:speciesrichness_data}
\end{figure}

For both groups of names, we then plot the neutral progeny distribution with maximum likelihood parameters alongside the empirical progeny distribution in Fig.~\ref{fig:speciesrichness_data}.  It is obvious that the neutral distribution (grey lines) produces too many
names with intermediate numbers of progeny relative to singletons (i.e. names
that have never been transmitted and therefore have an abundance of 1), and too few variants with very large numbers of progeny.

Given this discrepancy we ask whether novelty bias can provide a better explanation. Any form of pro-novelty bias, however, will only increase the differences (cf. Fig. \ref{fig:comp_neutral_novelty}) and therefore we focus on anti-novelty bias. 
Fig.~\ref{fig:speciesrichness_data} (red lines) shows the best fit over a discrete set of parameter values to the data. 
In order to replicate that only a relatively small (at least compared to the neutral predictions) number of innovations are transmitted at least once, we needed to choose $\ep_0=-1$ in Eq.~\eqref{eq:antiproglandscape}, so that new variants (initially) have zero competitive effect on any extant variant.  We also chose $\lambda >> b$, so that if a variant survives (meaning is transmitted at least once), it quickly begins to interact neutrally with the rest of the population. We note that we are not seeking to rigorously fit the anti-novelty bias model, but it is apparent that with these choices anti-novelty bias provides a potential explanation for the phenomena we see in these data.

\subsection{Restricting to Popular Names}

Our example data set above contains every baby name registered over a $70$-year period in a single region, leading to the potential conclusion that new, rare variants have a disadvantage. However, many available data sets for registered baby names in other regions are incomplete; providing only the most popular names due to privacy considerations. Previous studies have often tested hypotheses for cultural evolution based on similarly incomplete data and in this section we explore how this incompleteness may alter conclusions about the existence of selection biases in the population.

In the following we consider two common ways of preprocessing cultural frequency data, both of which amount to removing some subset of data. First, we only keep the most popular names over a given time span, removing any names with fewer appearances (in total, throughout the time interval) than a given threshold.  Second, we remove any names with less than a given threshold in any given year.  

\begin{figure}[htb]
\setlength{\unitlength}{0.1\textwidth}
\begin{center}
\includegraphics[angle=0,scale=0.175,trim= 0 0 0 0,clip=TRUE]{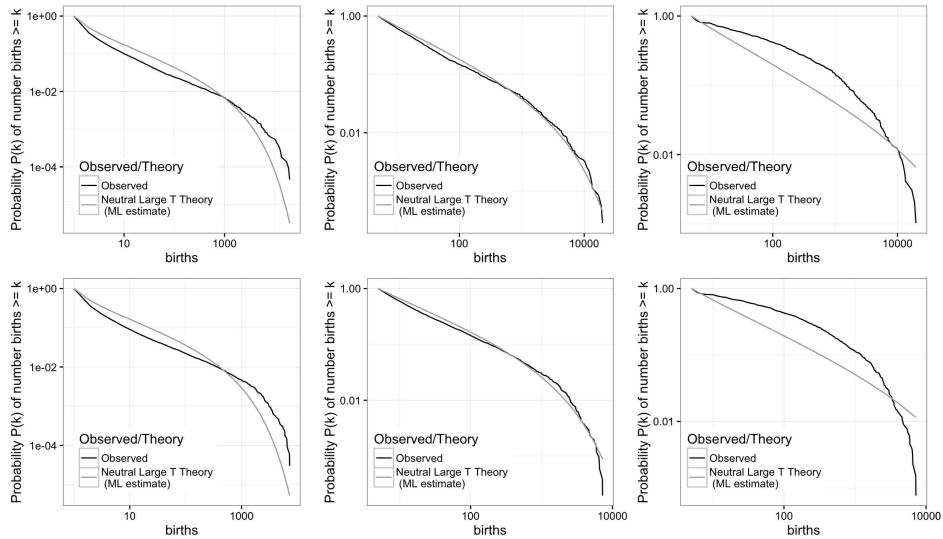}
\end{center}
\caption{Empirical progeny distribution and  maximum likelihood neutral progeny distribution generated by (left column) the full data set as in Fig.~\ref{fig:speciesrichness_data}, (middle column) imposing a threshold such that only names with five or more appearances are considered, (right column) by imposing a year-by-year threshold, such that only names with five or more appearances in a given year are considered in that year. Top row: boys names, bottom row: girls names. }\label{fig:speciesrichness_thresholding}
\end{figure}

In Figure~\ref{fig:speciesrichness_thresholding} we show the results of three analyses of the South Australia baby name data set (top row: boys names, bottom row: girls names).  Alongside our analysis using the full data set (left column), we also (i) remove names containing $<5$ instances over the $70$-year time span (middle column) and (ii) remove names from a given year that have $<5$ instances in that year (right column).  We call these a total threshold and a year-by-year threshold, respectively.  The differences between the three approaches are stark.  

We have seen in Section~\ref{sec:baby_names}(\ref{subsec:sa_analysis}) that the full progeny distribution can be replicated by assuming that innovations are strongly selected against but that this disadvantage fades away quickly; as soon as those novel names are transmitted. They then interact neutrally with the population and therefore we might expect that imposing the total threshold (i.e. in this case innovations are names whose progeny count exceeds this threshold) generates a distribution which is consistent with neutrality.  
However, if we impose the year-by-year threshold, the resulting progeny distribution changes substantially---if we treat this data as if all names were present, it would look consistent with a novelty advantage, rather than neutrality or novelty disadvantage.  The effect of these pre-processings of names data, and the qualitative differences they make, demonstrate the need to be cautious about any conclusions drawn using incomplete data.  Our results here mirror a long-standing debate in ecology on snapshots of species abundances, where a lack of sampling of rare species introduces what has been termed a `veil line', and can alter the shape of the species abundance distribution~\citep{preston1948commonness,nee1991lifting}. In our case, the progeny distribution veil line can lead us to infer a purely neutral explanation, where in reality there is a strong bias against new names.

\section{Discussion}

Innovation is ubiquitous across biological and social domains, but in many cases we lack a direct way to characterize the mechanisms of the innovation process. This is particularly true in the realm of cultural evolution, where it is often not obvious what to look for or to measure in a new variant to describe the mechanism  that gave rise to it. For example, the baby names considered in this paper have no direct analogue of beak size, body plan, or carbon fixation pathways.  
Nevertheless we know that in these domains new variants are `different' from  extant variants. 
In this paper we assumed that variants are functionally equivalent but differ in their ages and abundances in the population and aimed at understanding how theses differences can affect the spread behaviour of the innovations. To this end we analyzed the characteristics of the progeny distribution, which aggregates the temporal dynamics of new variants across the population over a fixed time interval, under different assumptions of cultural transmission. 

Using a mean field model drawn from ecology we derived the first analytical representation of the progeny distribution under the hypothesis of neutrality.
We showed that the neutral progeny distribution consists of two phases: a power law phase for intermediate numbers of progeny with a universally-applicable exponent of -3/2, followed by an exponential decay phase for large numbers of progeny. The onset of the exponential phase is modulated by the innovation rate: the higher the rate, the earlier the exponential cut-off. The analytical representation allowed us further to derive maximum likelihood estimates of the neutral model parameters, and therefore to establish whether observed empirical patterns are consistent with the hypothesis of neutrality. 

In order to allow for selective differences between the cultural variants 
we developed a simulation framework and analyzed the effects of pro-and anti-novelty biases on the shape of the progeny distribution. 
These biases alter the shape of the progeny distribution with pro-novelty biases increasing the occurrence of variants with a low or intermediate numbers of progeny and decreasing the occurrence of variants with a high numbers of progeny. These results go in hand with a decrease of the average life time of the individual variants. The reverse is true for anti-novelty bias. 

In applying our methodology to baby names from South Australia, we found that the data showed at least two different regimes. 
First, we see the generation of a lot of variation. The data sets contain a large number of innovations with abundance one, i.e. innovations that have never been transmitted. Second, we see the persistence of some names over a very long time. 
Our analysis showed that neutrality alone is not able to replicate these patterns, as it produces too many variants with intermediate numbers of progeny relative to singletons (i.e. names that have never been transmitted), and too few variants with very large numbers of progeny. The empirical progeny distribution of baby names is much more closely reflected by assuming an anti-novelty bias whereby the bias decays as soon as a variant survives long enough to become established. 
Importantly, we concluded that most new names do not proliferate, but if they are transmitted, their interactions with the other variants in the population quickly resemble those under neutrality  
(The code used for this analysis is available under\\ \url{https://github.com/odwyer-lab/neutral_progeny_distribution}).

This result points to the crucial importance of rare variants for reliable inference of processes of cultural evolution from aggregated population-level data in form of progeny distributions. Analyses based on incomplete data sets including only popular variants according to different threshold rules revealed consistency between the observed (incomplete) data and neutral evolution as well as pro-novelty bias.
This is a powerful reminder that we need to be cautious with conclusions about underlying evolutionary processes drawn from incomplete data. 

Lastly, we note that the result of this study is not to say that the choice of baby names {\it is} guided by anti-novelty bias but that anti-novelty bias is a potential cultural transmission process which could explain the observed, complete data set of baby names whereas neutral evolution and pro-novelty biases are not. There may be other, potentially more complex processes of cultural transmission which are able to replicate the observed progeny distribution equally well. For example, content bias might be producing a disadvantage for most new variants, leading to their early extinction, and leaving behind only those new variants which did not have this disadvantage. But the implication of this explanation is that content bias is fairly restrictive, with either a large negative, or neutral effect, but rarely (or never) a positive effect, a distribution which itself would require explanation. The extension of our analytical approach to incorporate these processes, alongside the inherent variability over time of real systems will help shading more light on this issue and be the focus of future research.


{\bf Acknowledgement.} JOD acknowledges the Simons Foundation Grant \#376199, McDonnell Foundation Grant \#220020439 and Templeton World Charity Foundation Grant \#TWCF0079/AB47. We thank the members of the Department of Human Behavior, Ecology and Culture at the Max Planck Institute for Evolutionary Anthropology for helpful comments on an earlier version of this manuscript. Further, we thank three anonymous reviewers for their helpful and encouraging comments.   

%
%
%


\setcounter{section}{0}
\setcounter{figure}{0}

\renewcommand{\thefigure}{S\arabic{figure}}
\renewcommand{\thetable}{S\arabic{table}}
\renewcommand{\thesection}{S\arabic{section}}
\renewcommand{\thesubsection}{S\arabic{section}.\arabic{subsection}}

\newpage

\begin{center}
{\bf \Large Supplementary material}
 \end{center}
   
 \section{Neutral Theory in Cultural Evolution}\label{sec:neutral_theory}

Neutral theory in cultural evolution has been mainly modelled using the Wright- Fisher infinitely many allele model (see e.g.~\citep{ewens2012mathematical} for a review of the mathematical properties,~\citep{neiman1995stylistic} for its introduction to cultural evolution as well as~\citep[e.g.][]{Shennan2001,kohler2004vessels,bentley2004random,steele2010ceramic} for further applications to cultural case studies).  
The theory assumes that in finite populations cultural variants are chosen to be copied according to their relative frequency, and new variants not previously seen in the populations are introduced by a process resembling random mutation. Changes in frequency therefore occur only as a result of drift. 
While these inherent assumptions are likely to be violated in the cultural context (for detailed discussions see~\citep[e.g.][]{neiman1995stylistic,Shennan2001,steele2010ceramic}) population-level patterns of various observed episodes of cultural change nevertheless resemble the ones expected under neutrality~\citep[e.g.][]{neiman1995stylistic,bentley2004random,bentley2007regular}. 
Importantly, theses studies do {\it not} conclude that neutral evolution is the underlying evolutionary force shaping the observed empirical patterns. They rather suggest that if each act of choosing one cultural variant rather than another has a different motivation, the emerging  population-level patterns will be that there are no directional selective forces affecting what is copied~\citep{shennan2011descent}.
However, it still has to be shown that neutral predictions are distinguishable from predictions generated by alternative cultural selection scenarios \citep[see e.g.][for a discussion of this issue in the ecological context]{rosindell2012case}. If a (potentially unknown) number of cultural scenarios result in very similar predictions, then the meaning of the rejection of the neutral hypothesis becomes hard to interpret. 

In the following we provide a brief overview over the characteristics of the Wright- Fisher infinitely many allele model. This model assumes that the composition of the population of instances of cultural variants at time $t$ is derived by sampling with replacement from the population of instances at time $t-1$ resulting in non- overlapping generations.  
The (temporally constant) population size $J$ and the variables $m_i$ and $n_i$ stand for the abundances of variant $i$ in the population at times $t-1$ and $t$, respectively. Then 
\[
p_i=\frac{m_i}{J}(1-\mu),\ i=1,2,\ldots
\]
describes the probability that a specific instance is of variant $i$. Further, $\mu$ denotes the innovation rate which describes the probability that a novel variant, not currently or previously seen in the population, is introduced. In general, the probability that the configuration of abundances $[m_1,m_2,\ldots]$ at time $t-1$ is transformed into $[n_0,n_1,n_2,\ldots]$ at  time $t$ is given by 
\begin{equation}\label{eq:infiniteAlleleModel}
P(X_0(t)=n_0,X_1(t)=n_1,\ldots|X_1(t-1)=m_1,\ldots)=\frac{J!}{\prod \limits_i m_i!}\prod\limits_i p_i^{n_i}
\end{equation}
with $p_0=\mu$ and $\sum_im_i =\sum_in_i=J$.  
The state space of the Markov process defined by these transition probabilities is extremely large making the derivation of population-level properties of this stochastic process almost intractable.
But Eq.~\eqref{eq:infiniteAlleleModel}  implies that the extinction of any variant is inevitable over time  and the time evolution of a single variant can be described by a  two-variant formulation 
\begin{equation}\label{eq:2AlleleModel}
P(X_i(t)=n_i|X_i(t-1)=m_i)=\binom{J}{n_i}p_i^{n_i}(1-p_i)^{J-n_i}.
\end{equation}
We note that under neutrality all variants are considered identical and therefore we can drop the index $i$ from Eq.~\eqref{eq:2AlleleModel}.

It follows from the Eq. \eqref{eq:2AlleleModel} that the probability that a newly introduced variant with abundance $1$ goes immediately extinct is given by
\[
P(X(t)=0|X(t-1)=1)=\left(1-\frac{1}{J}(1-\mu)\right)^J\rightarrow e^{\mu-1}\ \text{for large}\ J.
\]
Further, the diffusion approximation of Eq.\eqref{eq:2AlleleModel} allows us to determine  the transition probability density $f(x,p,\tau)$ as the solution of the diffusion equation 
\[
\frac{\partial f(x,p,\tau)}{\partial \tau}=a(p)\frac{\partial f(x,p,\tau)}{\partial p}+\frac{1}{2}b(p)\frac{\partial^2 f(x,p,\tau)}{\partial p^2}
\]
with $a(p)=-J\mu p$, $b(p)=p(1-p)$ and appropriately scaled space and time dimensions $p=m/J$, $x=n/J$ and $\tau=t/J$~\citep[e.g.][]{kimura1964diffusion}. In general, an explicit solutions of this equation can only be achieved under relatively restrictive assumptions,~\citep[e.g for $\mu=0$,][]{kimura1955random}.
Nevertheless, it has been shown that some steady-state properties of the population of instances of cultural variants can be determined. 
The variant abundance distribution describing the expected number of variants with relative frequencies in the interval $(x,x+\delta x)$ at steady state can be approximated by 
\begin{equation}\label{eq:variantAbundanceWF}
\phi(x)=\theta_c x^{-1}{(1-x)^{\theta_c-1}}
\end{equation}
with $\theta_c=2J\mu$~\citep{kimura1964number}.
Additionally, the average number of different variants, $S$, in the populations can be described by
\[
E\{S\}=\theta_c+\int\limits_{1/J}^1 \theta_cx^{-1}{(1-x)^{\theta_c-1}}dx
\]
(e.g.~\citep{ewens2012mathematical}). 

We note that the variant abundance distributions given by Eq. \eqref{eq:speciesAbundanceNSZ} in the main text and Eq. \eqref{eq:variantAbundanceWF} generate similar results for sufficiently large $J$ and sufficiently small $\nu$.



\subsection{Simulation of the Wright-Fisher model}
\label{sec:simwf}

Simulations of the infinitely many allele Wright-Fisher model are relatively easily obtained through random sampling from previous generations. In detail, in each time step $t$ a new set of $J$ instances is generated through random copying from the population of instances of cultural variants at time step $t-1$ possessing the abundance configuration $[m_1,m_2,\ldots,m_{S(t-1)}]$ with $\sum\limits_{i=1}^{S_{t-1}}m_i=J$. The variable $S_{t-1}$ stands for the number of different variants at time step $t-1$ and $m_i$ records their abundance.
The probability that variant $i$ is copied in each of the $J$ production events is given by $p_i=\frac{m_i}{J}(1-\mu)$ where $\mu$ stands to the innovation rate. If an innovation occurs then a variant, not currently or previously seen in the population, is introduced. 

\begin{figure}[htb]
\vspace*{0.3cm}
\begin{center}
\includegraphics[width=0.6\textwidth]{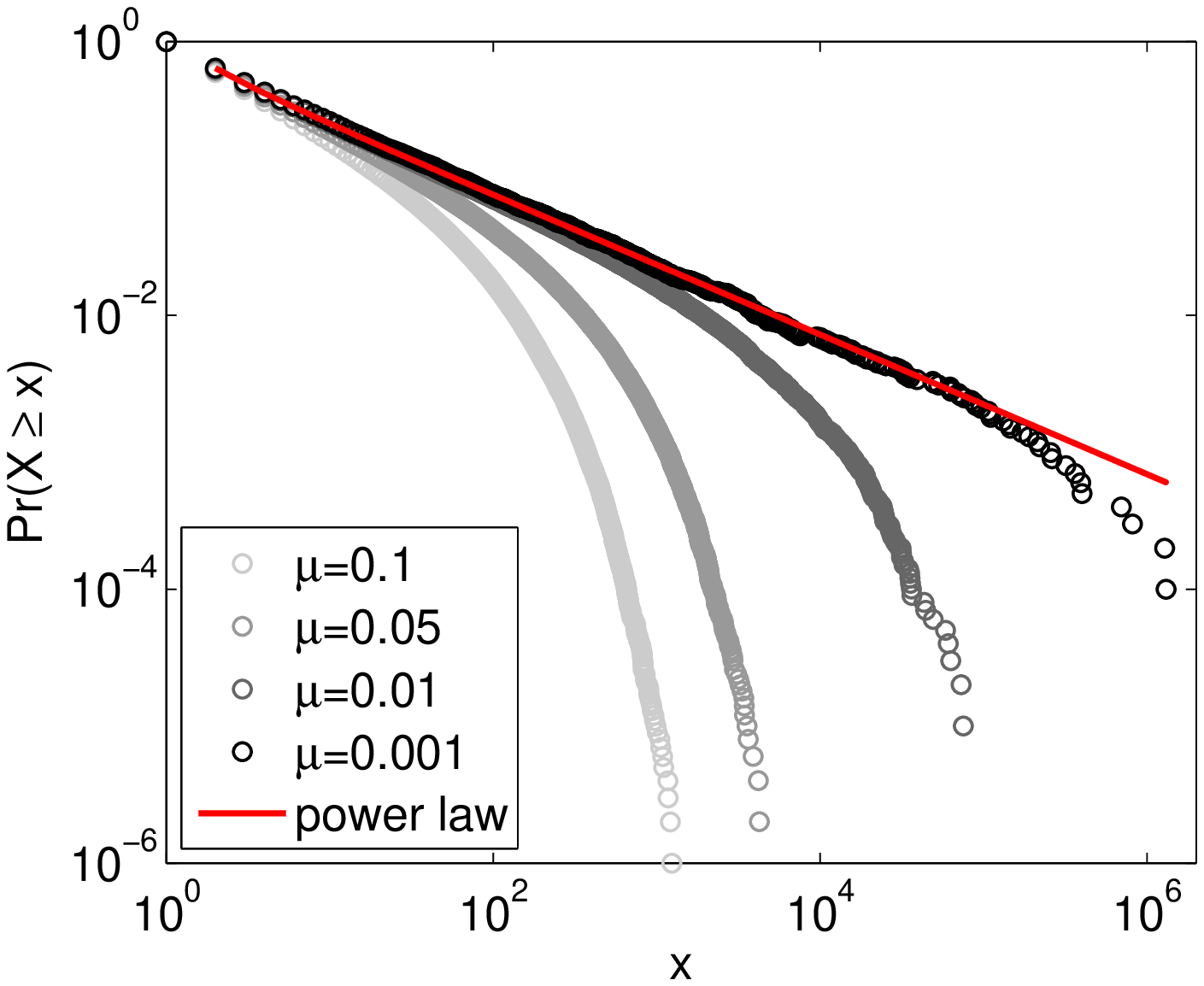}
\end{center}
\caption{Progeny distribution determined by the Wright-Fisher infinitely-many alleles model for $J=1000$, $T=200,000$ and $\mu=0.001$ (black circles), 0.01 (dark gray circles), 0.05 (gray circles), 0.1 (light gray circles).} 
\label{fig:DBN_WF}
\end{figure}

After a burn in period which ensures that the system has reached an approximate steady state we determine the progeny distribution after $T=200,000$ generation for $J=1,000$ and various values of $\mu$ (see dotted lines in Fig. \ref{fig:DBN_WF}).  
We lack an analytical result for the cumulative Wright-Fisher progeny distribution, but drawing from our results for the overlapping generations neutral model we plot a power law with exponent $-1/2$ (red line) (we showed in the main text that for intermediate values of $k$ the progeny distribution resembles a power law with exponent $-3/2$). As $\mu$ becomes small, we can see that this power law with fixed exponent becomes an increasingly accurate explanation of the first phase of this distribution, just as in the case of overlapping generations. It is likely that fitting a single power law to the whole distribution, including the exponential decline, would explain the apparent variation in power law exponent with $\mu$ and $J$ identified in earlier studies.

\section{NZS Solutions for Species Dynamics and Species Abundance Distribution}\label{sec:NZSSAD}

The non-zero sum (NZS) formulation of neutral theory is an approximation to a neutral, overlapping generations model where all variants compete for a single resource, and the strength of competitive interactions is equal across all pairs of variants.  The defining master equation focuses on the dynamics of one focal variant, and characterizes its change in abundance through time, from an initial condition (usually taken to be $n=1$, and known as point speciation in the ecology literature)
\begin{equation}\label{eq:master}
\frac{dP}{dt} = b(n-1)P(n-1|t)-bnP(n|t)-dnP(n|t)+d(n+1)P(n+1|t).
\end{equation}
This master equation is linear because the interactions between the focal variant and the rest of the population are treated in a mean field approximation. In effect, this equation assumes that the remainder of the population is of constant size, and then the pairwise competitive interactions are approximated by just adding to the mortality rate for this variant.

To solve Eq.~\eqref{eq:master} for $P(n|t)$, we use the generating function $G(z,t)$ defined by 
\begin{equation*}
G(z,t) = \sum_k P(n|t)z^k
\end{equation*}
which in turn is the solution of
\begin{align*}
\frac{\partial G}{\partial t} = & (z-1)\lt(b(z-1)-(d-b)\rt)\frac{\partial G}{\partial z}.
\end{align*}
Using the method of characteristics, it can be shown that the equation above is solved by 
\begin{equation}\label{eq:sol_G}
G(z,t) = 1+\frac{e^{-\nu t} (z-1)}{1- \frac{b}{\nu}(1-e^{-\nu t})(z-1)}.
\end{equation}
Consistent with the main text, the speciation rate is defined by $\nu=d-b$. To obtain the solution~\eqref{eq:sol_G} we imposed $G(1,t)=1$ ensuring the normalization of the probability distribution $P(n|t)$ (i.e. the sum over all values of $n$ is equal to one), and $G(z,0)=z$ corresponding to the point speciation initial condition $n=1$. 

Eq.~\eqref{eq:sol_G} is the generating function of an exponential distribution with time-varying coefficients and the explicit solution of Eq.~\eqref{eq:master}  is therefore obtained by transforming back from this generating function to the exponential $P(n|t)$.  For $n\ge 1$, it holds 
\begin{align}
P(n|t) & = \frac{e^{-\nu t}}{(\nu+b(1-e^{-\nu t}))^2}  \lt[\frac{b(1-e^{-\nu t})}{\nu+b(1-e^{-\nu t})}\rt]^{n-1}~\label{eq:solution},
\end{align}
while for $n=0$
\begin{equation*}
P(0|t)  =  1-\frac{e^{-\nu t} }{1+ \frac{b}{\nu}(1-e^{-\nu t})}.
\end{equation*}
The expected species richness in this model is given by
\begin{align*}
S& = \nu J \sum_{n=1}^{\infty}\int_{0}^{\infty} dt P(n|t)\nn\\
& = \nu J \int dt \frac{e^{-\nu t} }{1+ \frac{b}{\nu}(1-e^{-\nu t})}\nn\\
& = \frac{\nu J}{b} \log\lt(\frac{b}{\nu}\rt),
\end{align*}
i.e. we sum over all speciation events in the history of the population (of total size $J$), and compute the probability of those variants being in the population in the present time.  Similarly, the expected  distribution of variant abundances (known as the species abundance distribution in the ecology literature) in this model is given by
\begin{align*}
S(n) & = \nu J \int_{0}^{\infty} dt P(n|t)\nn\\
& = \nu J \int_{0}^{\infty} dt\  \frac{e^{-\nu t}}{(\nu+b(1-e^{-\nu t}))^2}  \lt[\frac{b(1-e^{-\nu t})}{\nu+b(1-e^{-\nu t})}\rt]^{n-1}\nn\\
& = \frac{\nu J}{b} \lt[\frac{b}{b+\nu}\rt]^n.
\end{align*}

\section{NZS Solution for the Progeny Distribution}\label{sec:NZSPD}

We now derive the joint probability distribution $Q(n,k|T,n_0)$ that after time $T$, a variant has $n$ extant individuals, and has had a total of $k$ birth events during the time interval from $0$ to $T$, conditioned on the initial abundance $n_0$ at time $0$. Marginalizing $Q(n,k|T,n_0)$ will lead to a prediction for the neutral progeny distribution, a quantity rarely considered in ecological contexts, but used as a test of neutrality in cultural evolution.  Note that we are not necessarily starting this time interval at the speciation time, and so the variant could have some arbitrary abundance $n_0$ at the start of our time interval. Initially, though, we will drop the $n_0$-dependence and work with  initial condition $n_0=1$.  

For the birth death process described in the last section it holds
\begin{eqnarray*}
\frac{dQ}{dT} = && b(n-1)Q(n-1,k-1|T)-bnQ(n,k|T)\\
&+& d(n+1)Q(n+1,k|T)-dnQ(n,k|T).
\end{eqnarray*}
Note that $k$ does not affect any of the rates. We now consider a new generating function, $G(z,y,T)$, defined as
\begin{equation*}
G(z,y,T) = \sum_{n=0}^{\infty}\sum_{k=0}^{\infty}Q(n,k|T)z^ny^k
\end{equation*}
which then satisfies
\begin{equation}\label{eq:G}
\frac{\partial G}{\partial T} = [bz(yz-1)-d(z-1)]\frac{\partial G}{\partial z}
\end{equation}
with initial and boundary conditions
\begin{eqnarray}\label{eq:iniCond}
G(1,1,T) &=& 1,\nonumber\\
G(z,y,0) &=& z 
\end{eqnarray}
For a more general initial condition $n_0\neq 1$ the latter condition changes to $z^{n_0}$.

Eq. \eqref{eq:G} has a solution of the form 
\begin{align*}
G(z,y,T) = \frac{A(y)-C(y)\lt(\frac{A(y)-B(y)z}{C(y)+B(y)z}\rt)e^{T/F(y)}}{B(y)\lt[\lt(\frac{A(y)-B(y)z}{C(y)+B(y)z}\rt)e^{T/F(y)}+1\rt]}
\end{align*}
with
\begin{align*}
F(y) & = \lt[(b+d)^2-4bdy\rt]^{-\frac{1}{2}},\nn\\
A(y)&  = 1+F(y)(b+d),\nn\\
B(y)& = 2byF(y),\nn\\
C(y)& =1-F(y)(b+d).
\end{align*}
Due to the linear nature of the problem the  solution for more general initial conditions, $n_0$, is given by 
\begin{equation*}
G(z,y,T,n_0)=G(z,y,T)^{n_0}.
\end{equation*}
Finally, we can marginalize over the unobserved $n$ (assuming we have knowledge about the progeny, and not about total abundances/census counts) by setting $z=1$
\begin{align*}
H(y,T,n_0)&  = G(1,y,T)^{n_0}\nn\\
& = \lt( \frac{A(y)-C(y)\lt(\frac{A(y)-B(y)}{C(y)+B(y)}\rt)e^{T/F(y)}}{B(y)\lt[\lt(\frac{A(y)-B(y)}{C(y)+B(y)}\rt)e^{T/F(y)}+1\rt]}\rt)^{n_0}.
\end{align*}
Weighting $q(k|T,n_0)$ by the steady state species abundance distribution and taking the asymptotic limit of large $T$ leads to
\begin{align*}
H_{\textrm{extant}}(y,T)& = \sum_{n_0} S(n_0) H(y,T ,n_0)\nn\\
 =&  \sum_{n_0} S(n_0)\lt( \frac{A(y)-C(y)\lt(\frac{A(y)-B(y)}{C(y)+B(y)}\rt)e^{T/F(y)}}{B(y)\lt[\lt(\frac{A(y)-B(y)}{C(y)+B(y)}\rt)e^{T/F(y)}+1\rt]}\rt)^{n_0}\nn\\
=&  -\frac{\nu J}{b} \log\lt[1-\frac{b}{d}\lt( \frac{A(y)-C(y)\lt(\frac{A(y)-B(y)}{C(y)+B(y)}\rt)e^{T/F(y)}}{B(y)\lt[\lt(\frac{A(y)-B(y)}{C(y)+B(y)}\rt)e^{T/F(y)}+1\rt]}\rt)\rt].
\end{align*}

We do not yet account for new variants that can appear during the interval $T$, and themselves contribute to this birth event count. 
To include these instances we change the initial condition \eqref{eq:iniCond} to $G(z,y,0)=y$, 
i.e. there is one instance in both, the variant population and its progeny distribution, immediately at speciation. Therefore, this contribution takes the form
\begin{equation*}
\nu J\int_0^T d\tau y H(y,\tau,1)
\end{equation*}
with an extra factor of $y$ relative to the results above. This means new variants arise at a rate $\nu J$ per unit time, they begin per definition with a single instance and single contribution to the progeny distribution, and persist from their innovation time up until $T$.  So in total
\begin{eqnarray}\label{eq:fullcumuldist}
H_{\textrm{total}}(y,T) &=& H_{\textrm{extant}}(y,T)+H_{\textrm{new}}(y,T)\\
& = &-\frac{\nu J}{b} \log\lt[1-\frac{b}{d}\lt( \frac{A(y)-C(y)\lt(\frac{A(y)-B(y)}{C(y)+B(y)}\rt)e^{T/F(y)}}{B(y)\lt[\lt(\frac{A(y)-B(y)}{C(y)+B(y)}\rt)e^{T/F(y)}+1\rt]}\rt)\rt]\nonumber\\
&& \hspace*{-1.0cm} +\nu J\lt(y\frac{A(y)}{B(y)}T-\frac{2yF(y)}{B(y)}\log\lt[\frac{C(y)+B(y)+(A(y)-B(y))e^{T/F(y)}}{2}\rt]\rt).\nonumber
\end{eqnarray}
This is the generating function of the neutral progeny distribution, under the non-zero sum formulation of the neutral theory.

\subsection{Approximations for large $T$}

For large $T$, it holds 
\begin{align*}
H(y,T,n_0)& \simeq \lt(-\frac{C(y)}{B(y)}\rt)^{n_0} =\lt(-\frac{1 - F(y) (b + d)} {2 byF(y)}\rt)^{n_0}.
\end{align*}
Keeping only this leading term of this expansion, and considering the special case of $n_0=1$, $H(y,T,n_0)$ can be inverted analytically to give
\begin{equation}\label{eq:q}
q(k|T,1) \simeq \frac{2d}{b+d}\lt(\frac{4bd}{(b+d)^2}\rt)^k (-1)^k {{\frac{1}{2}}\choose{1+k} }.
\end{equation}
There is a power law phase $\propto k^{-3/2}$ resulting from the asymptotics of the binomial coefficient, and for sufficiently large $k$ there is an exponential drop-off.   Eq. \eqref{eq:q} can be written in terms of the per capita speciation rate $\nu$, as 
\begin{align*}
q(k|T,1) & \simeq \frac{2}{2-\frac{\nu}{d}}\lt(\frac{4(1-\frac{\nu}{d})}{4(1-\frac{\nu}{d})+\lt(\frac{\nu}{d}\rt)^2}\rt)^k (-1)^k {{\frac{1}{2}}\choose{1+k} }\nn\\
& = \frac{2}{2-\frac{\nu}{d}}\lt(1+\frac{\lt(\frac{\nu}{d}\rt)^2}{4(1-\nu/d)}\rt)^{-k} (-1)^k {{\frac{1}{2}}\choose{1+k} }.
\end{align*}
For small enough $\nu$ the exponential phase kicks in only for relatively large cumulative abundances, i.e. for small $\nu$, it holds 
\begin{align*}
q(k|T,1) & \simeq \lt(1+\frac{\nu}{2d}\rt)e^{-\lt(\frac{\nu}{2d}\rt)^2k}\ (-1)^k {{\frac{1}{2}}\choose{1+k} }
\end{align*}
which could be compared to the $\nu$ dependence of the species abundance distribution $S(n)$. 

\smallskip

This concludes the consideration of a single variant, with $n_0=1$ instances initially.  Because each variants is guaranteed to go extinct ($d>b$ in the NZS neutral model), there is a finite solution for the cumulative birth distribution at late times. If we now turn to the whole population,  represented by $H_{\textrm{total}}(y,T) $, we encounter a problem.  The first term $H_{\textrm{extant}}(y,T)$ is finite, as all of the variants summed over will go extinct and produce a finite number of birth counts.  However, the second term  $H_{\textrm{new}}(y,T)$ will produce an infinite number of birth counts, and eventually will dwarf the contribution from the steady-state variants contained in $h_{\textrm{total}}(y)$, i.e will dwarf contributions from variants that were already present at $T=0$.  Consequently, if the population persists indefinitely, all those initial variants will produce their contribution to the birth counts and eventually die out. The population, however, will continue to exist via new variants and the limit for the total number of births will tend to $\infty$.

\smallskip

We start with examining the limit of large $T$ for $H_{\textrm{extant}}(y,T)$
\begin{eqnarray}\label{eq:limHextant}
 \lim_{T\rightarrow\infty}H_{\textrm{extant}}(y,T)& = &\lim_{T\rightarrow\infty} \sum_{n_0} S(n_0) H(y,T,n_0)\\
 &=& \sum_{n_0} S(n_0)\lt(-\frac{1 - F(y) (b + d)} {2 byF(y)}\rt)^{n_0}\nonumber \\
&=& -\frac{\nu J}{b} \log\lt[\frac{-(b + d) + 2 d y + \sqrt{(b + d)^2 - 4 b d y})}{2 d y}\rt].\nonumber
\end{eqnarray}
There is no analytical expression for the distribution corresponding to this generating function, i.e. Eq \eqref{eq:limHextant} cannot be inverted analytically. But using numerical techniques we confirm that the generating function produces a distribution characterized by a power law with exponent $-3/2$ followed by exponential decline.  

Further, for large $T$, it holds for the new variants
\begin{equation}\label{eq:Hnew}
H_{\textrm{new}}(y,T)\simeq -\nu J Ty\frac{C(y)}{B(y)} =  -\nu J T\frac{1 - F(y) (b + d)} {2 bF(y)}.
\end{equation}
As pointed out above, there are an unbounded number of birth events from new variants introduced during the interval $T$, and expression \eqref{eq:Hnew} (valid for large $T$) will eventually dominate the finite numbers coming from the term $H_{\textrm{extant}}(y,T)$. The total number of births from new and extant variants are equal when roughly $T\sim\frac{1}{\nu}$. Beyond this point there are very few instances from the extant variants at $T=0$, and an ever increasing number from novel variants introduced during the considered interval.  Note also that this is not a normalized distribution yet and therefore it is not problematic that its coefficients diverge for large $T$: the coefficients of this generating function are the actual number of variants producing a given cumulative number of births, not the probability that a single variant will produce a given number of births.  However, normalization leads to 
\begin{align*}
\frac{H_{\textrm{total}}(y,T)}{H_{\textrm{total}}(1,T)} & \simeq \frac{yC(y)/B(y)}{C(1)/B(1)} = \frac{yC(y)}{B(y)}\frac{2b}{(d-b)\lt(1-\frac{b+d}{d-b}\rt)}\nn\\
& = -\frac{yC(y)}{B(y)}
\end{align*}
for late times $T$. This normalized distribution at very late times is given exactly analytically by the same distribution we found above, but with $k\rightarrow k-1$ reflecting the fact that the initial single instance already counts as a birth event.  So it always holds $k>0$ and we obtain
\begin{equation}\label{eq:late-time-SI}
q(k)= (-1)^{k-1}k {{\frac{1}{2}}\choose{k} } \frac{2d}{b+d}\lt(\frac{4bd}{(b+d)^2}\rt)^{k-1}  .
\end{equation}
This of course comes from the fact that at large enough $T$, we are just summing together the entire number of births for multiple variants starting with $n_0=1$. The corresponding cumulative distribution is straightforward to compute analytically in terms of a hypergeometric function for the cumulative distribution for \eqref{eq:late-time-SI}. Putting it together leads to 
\begin{equation*}
\hspace*{-0.25cm}
P(K\ge k) =  (-1)^{k-1}\frac{b+d}{2b}\lt(\frac{4bd}{(b+d)^2}\rt)^k {{\frac{1}{2}}\choose{k} }\pFq{2}{1}{1,(-1/2+k)}{1+k}{\frac{4bd}{(b+d)^2}}.
\end{equation*}

\smallskip

\section{Maximum likelihood estimation}\label{App:MaxLik}

In this section we derive the maximum likelihood estimate of the ratio $\eta=d/b$.

The log likelihood of observing a given set of $S$ cultural variants with abundances $\{k_i\}$ at late times is given by
\begin{equation*}
L = \sum_{i=1}^S log(q(k_i)) = \sum_{i=1}^S    \log\lt[\frac{2d}{b+d}\lt(\frac{4bd}{(b+d)^2}\rt)^{k_i-1} (-1)^{k_i-1} {{\frac{1}{2}}\choose{k_i} }\rt]
\end{equation*}
which can be rewritten as
\begin{equation*}
L =\sum_{i=1}^S   \log\lt[\frac{2\eta}{1+\eta}\lt(\frac{4}{\frac{1}{\eta}+2+\eta}\rt)^{k_i-1} (-1)^{k_i-1} {{\frac{1}{2}}\choose{k_i} }\rt]
\end{equation*}
by using the relation
$\eta= \frac{d}{b}=\frac{\nu}{b}+1$.
It holds 
\begin{equation*}
\frac{\partial L}{\partial \eta}=\sum_{i=1}^S  (k_i-1)\frac{\eta-1}{\eta(1+\eta)}+ \frac{S}{\eta(1+\eta)}.
\end{equation*}
Setting $K_{\text{total}}=\sum_{i=1}^S  k_i$ and solving $\frac{\partial L}{\partial \eta}=0$ leads to
\[
\eta=\frac{K_{\text{total}}}{K_{\text{total}}-S}.
\]


\section{Simulation of the Overlapping Generations Model via Gillespie algorithm}
\label{sec:simoverlap}

The NZS approximation described in section~2(\ref{sec:neutral_ecology}) in the main text has been extensively compared with both, simulations and analytical results for ecological populations with symmetric, competitive interactions.
In general, it has been demonstrated that the predictions of the NZS approximation for the distribution of variant abundances at a single point are valid when the innovation rate satisfies $\nu J>>1$, and begin to break down when  $\nu J$ is small. To test the validity of the approximation~\eqref{eq:late-time} given in the main text, we take the same approach and simulate a group of competing variants, but compute the resulting progeny distribution after a long time interval, rather than the species abundance distribution at a single point in time. 

The simulated populations are described by stochastic Lotka-Volterra systems, where variant $i$ with current abundance $n_i$ will increase abundance by one individual at a rate $b_0n_i$, undergo intrinsic mortality and decrease abundance by one at a rate $d_0n_i$. 
Further, competitive interactions involve the focal variant of abundance $n_i$ in a population of current size $J$ and occur at a rate $\alpha n_i J$.  The strength of competition is controlled by the parameter $\alpha$ and its outcome is the loss of one instance either from the focal variant or from the rest of the population. 
New variants are introduced at a rate $\nu J$ with initial abundance $1$, and are considered as an error in the birth process. Therefore the effective per capita birth rate (i.e. the rate of production of instances of the same variant) is $b_0-\nu$.  In summary, the rates of these processes for variant $i$ are as follows
\begin{align}\label{eq:summary_processes_rates}
\hspace{0.1cm}\textrm{process} \hspace{1cm}& \hspace{1cm} \textrm{description} \hspace{0.5cm} & \hspace{0.1cm} \textrm{rate}\nn\\
\hspace{0.1cm} n_i \rightarrow n_i +1\hspace{1cm} &\hspace{1cm} \textrm{birth} \hspace{0.5cm}&\hspace{0.1cm} (b_0-\nu)n_i\nn\\
\hspace{0.1cm}n_i \rightarrow n_i -1\hspace{1cm} & \hspace{1cm}\textrm{intrinsic mortality}\hspace{0.5cm} & \hspace{0.1cm} d_0n_i\nn\\
\hspace{0.1cm}n_i \rightarrow n_i -1\hspace{1cm} &\hspace{1cm} \textrm{competition}\hspace{0.5cm} &\hspace{0.1cm} \alpha n_i \sum_{\forall\,j} n_j\nn\\
\hspace{0.1cm}0 \rightarrow 1 \hspace{1cm} &\hspace{1cm} \textrm{speciation}\hspace{0.5cm} &\hspace{0.1cm} \nu\sum_{\forall\,j} n_j
\end{align}
where the labels $i$ and $j$ refer to the extant variants in the system at any given point in time, and the sums are taken over all variants, including variant type $i$.

The simulation of this population is based on the well-known Gillespie algorithm~\citep{gillespie1976general}.  This approach involves a sequence of transitions drawn from the possibilities given in \eqref{eq:summary_processes_rates}, with a waiting time in between each of these events.  For example, for a system with three variants, a birth event for one of the three types could be followed by a competitive interaction between the other two with the outcome that type three loses an instance, and so on. For a given configuration of instances the waiting time between two events is distributed according to an exponential distribution with a mean time equal to the inverse of the sum of all rates.  
When an event occurs, the kind of transition is randomly chosen with weights proportional to their rates.
Consequently, events are more likely to involve an abundant variants, because all processes are weighted by total variant abundance (see \eqref{eq:summary_processes_rates}).  

Finally, the intrinsic rates given in \eqref{eq:summary_processes_rates} represent the exact description of the population dynamics. In order to evaluate the accuracy of the NZS approximation we need to
map those intrinsic rates onto the parameters of the NZS approximation.
This mapping is such that the \textit{effective} birth rate of each variant is given by $b=b_0-\nu$, while the effective mortality rate (incorporating both intrinsic mortality and competition) is given by $d=b_0$. As $d_0$ does not directly enter the NZS prediction for the progeny distribution, we simulated these populations with $d_0=0$.

The NSZ expectation for the steady-state population size was derived in~\citep{odwyer2014redqueen}
\begin{equation}
J_{\textrm{steady}} = \frac{b_0 - d_0}{\alpha}
\end{equation}
which simplifies to $b_0/\alpha$ when the intrinsic mortality vanishes.  We therefore set an initial condition for the simulated population of $b_0/\alpha$ instances of only one cultural variant.  

To ensure that the system has reached an approximate steady-state before we begin sampling the progeny distribution, we allow the system to burn in by waiting until the first monodominant variant has reached extinction. At this point \textit{every extant variant has experienced entirely neutral dynamics, starting from a single instance}, and therefore no deviation from the average steady-state neutral population is expected.  From this point onwards, we record all birth events, and begin accumulating the progeny distribution.  
In order to provide a valid comparison with the late-time limit of the progeny distribution given by Eq. \eqref{eq:late-time} in the main text, we stop sampling after $T=100b_0/\nu$ time steps (see section~\ref{sec:NZSPD} for a derivation of this stopping time). Additionally, we verified that the first two moments of the progeny distribution were asymptotic to constant values by this time, and therefore ensured that we indeed sampled the asymptotic progeny distribution for large $T$.

\section{Data set}\label{app:data_set}

The South Australian Attorney-General's department provides two data sets consisting of all boys' and girls' names registered from 1944 to 2013, respectively, in South Australia. These data sets can be found under (last accessed 27.02.2017) \\
{\it https://data.sa.gov.au/data/dataset/popular-baby-names}

Between 1944-2013, the total number of girls  names registered each year varied from a low of $6748$ (in 1944) to a peak of $11754$ (in 1971), subsequently declining slightly to between $9000-10000$ in the last three decades. The total number of distinct names registered each year varied between $741$ and $2923$. For boys, the total number of names registered per year varied between $7069$ and $12464$, following a similar pattern to the girls' names, while the total number of distinct names registered each year varied between $477$ and $2450$. Clearly, there is systematic variation here in both numbers of names (reflecting a changing population size) and in the diversity of names (potentially reflecting a non-stationarity in the innovation rate). However, this variation may be as small as we can reasonably expect in cultural data. 

We also note that this $70$ year span is not a priori long enough to apply our asymptotic results. But we have also explored the change in the progeny distribution over time by considering the change in its first two moments as the time interval, $T$, over which the progeny distribution is computed is varied from one year up to $70$ years.  If these moments asymptote to a constant, this would indicate that the distribution is approaching its asymptotic form.  We find that these moments are still changing in time as $T$ approaches $70$ years, but that this change is relatively slow, indicating that this value of $T$ is close to the asymptotic regime.  
Therefore we propose that it is reasonable to apply our methodological approach, which assumes that the system is in a steady state with a temporally constant innovation rate $\nu$, 
and compare the Australian baby name data to the asymptotic form of the progeny distribution for large time intervals that we derived in Eq.~\eqref{eq:late-time-cumulative} in the main text.

We stress that in general we have to be careful in drawing conclusions from observed data too firmly. In part because the data likely does not reflect a population in steady-state, or with a constant innovation rate over time, and may only barely span a long enough time frame for our asymptotic results to be applicable.  But at the very least, our approach might lead to ways to incorporate this variation, which is inevitably present in real data, and has been underexplored in studies of cultural evolution so far.  

Additionally we note that different geographical regions will differ in their legislation towards the use of novel baby names (e.g. administrative approval processes might be more or less stringent) 
which naturally influences the rate of innovation. 
But our analysis is focused on the spread behavior of innovation, i.e. variants that have been introduced into the system with abundance one. Our results indicate that e.g. the ratio between singletons and variants with abundance two is sensitive to the underlying  process of cultural transmission. External processes affecting the rate of innovation might not influence this ratio strongly. 
Further, the size of the `name space' (meaning the space of all feasible names given the conventions of the particular language) is usually not known. This leads to the question whether the name space could become exhausted over time resulting in a decline of the innovation rate. While this is a valid concern we did not see a strong indication of such a phenomenon in the considered data set: the innovation rates did not show a strong decline over time.

\bibliographystyle{unsrtnat}
\bibliography{refs}{}


\end{document}